\documentclass[manuscript,sigconf]{acmart}

\usepackage{xfrac}
\usepackage{enumitem}
\usepackage{subcaption}
\usepackage{outlines}

\usepackage{xspace}
\usepackage{graphicx} 
\usepackage{grffile}
\newcommand{\blockfl}{\textsf{BlockFLA}\xspace}

\AtBeginDocument{%
  \providecommand\BibTeX{{%
    \normalfont B\kern-0.5em{\scshape i\kern-0.25em b}\kern-0.8em\TeX}}}

\newcommand{\eg}{e.g.\xspace}
\newcommand{\ie}{i.e.\xspace}

\newcommand{\ts}{\textsuperscript}

\DeclareMathOperator*{\argmin}{arg\,min}
\DeclareMathOperator*{\g}{\textbf{g}}
\DeclareMathOperator{\sign}{sgn}

\begin{document}
\title{\blockfl: Accountable Federated Learning \\ via Hybrid Blockchain Architecture}

\author{Harsh Bimal Desai}
\affiliation{%
  \institution{The University of Texas at Dallas}
  \city{Richardson}
  \state{Texas}
  \country{USA}}
\email{hbd140030@utdallas.edu}

\author{Mustafa Safa Ozdayi}
\affiliation{%
  \institution{The University of Texas at Dallas}
  \city{Richardson}
  \state{Texas}
  \country{USA}}
\email{mustafa.ozdayi@utdallas.edu}

\author{Murat Kantarcioglu}
\affiliation{%
  \institution{The University of Texas at Dallas}
  \city{Richardson}
  \state{Texas}
  \country{USA}}
\email{muratk@utdallas.edu}

\renewcommand{\shortauthors}{a1 and a2, et al.}

\begin{abstract} 

Federated Learning (FL) is a distributed, and decentralized machine learning protocol. By executing FL, a set of agents can jointly train a model without sharing their datasets with each other, or a third-party. This makes FL particularly suitable for settings where data privacy is desired.

At the same time, concealing training data gives attackers an opportunity to inject backdoors into the trained model. It has been shown that an attacker can inject backdoors to the trained model during FL, and then can leverage the backdoor to make the model misclassify later.  Several works tried to alleviate this threat by designing robust aggregation functions. However, given more sophisticated attacks are developed over time, which by-pass the existing defenses, we approach this problem from a complementary angle in this work. Particularly, we aim to discourage backdoor attacks by detecting, and punishing the attackers, possibly after the end of training phase.

To this end, we develop a hybrid blockchain-based FL framework that uses smart contracts to automatically detect, and punish the attackers via monetary penalties. Our framework is general in the sense that, any aggregation function, and any attacker detection algorithm can be plugged into it. We conduct experiments to demonstrate that our framework preserves the communication-efficient nature of FL, and provide empirical results to illustrate that it can successfully penalize attackers by leveraging our novel attacker detection algorithm.
\end{abstract}


\maketitle

\section{Introduction}\label{sec:intro}

Federated Learning~\cite{mcmahan2016communicationefficient} (FL) is a multi-round machine learning protocol that is run between an aggregation server and a set of agents. FL allows the participating agents to collaboratively train a model without sharing their data with each other, or with a third-party. At a high level, each agent first locally trains a model on his dataset, and then send his model to the server for aggregation. In return, the server aggregates the received models,
and returns the aggregated model back to agents for the next round of training. The rounds can simply go on until the trained model reaches some desired performance metric (e.g., accuracy) on a validation dataset maintained by the server. Since the data does not leave its owner, FL is particularly suitable for settings where privacy-sensitive data is involved. A vast range of organizations can collaborate on training a model via FL, and obtain a better performing model with respect to a model that is only trained on locally available data, while maintaining privacy of the data.

However, since the data of agents is unvetted, FL is susceptible to a wide range of attacks.
We particularly consider \emph{backdoor attacks}~\cite{arxiv:2018:fedlens,arxiv:2018:backdoor} in this work as they are the biggest threat for FL to the best of our knowledge. In a backdoor attack, an adversary disturbs the training process to make the model learn a \emph{targeted misclassification functionality}~\cite{chen2017targeted,shafahi2018poison,liu2017trojaning}. 
In centralized setting, this is typically done by \emph{data poisoning}. For example, in a classification task involving dogs and birds, the adversary could label all blue birds in the training data as dogs in an attempt to make the model to classify blue birds as dogs at the inference/test phase.  In FL, since the data is decentralized, it is unlikely that an adversary could access all the training data. Thus, backdoor attacks are typically carried through \emph{model poisoning} in the FL context ~\cite{arxiv:2018:fedlens,arxiv:2018:backdoor,sun2019really}. That is, the adversary tries to constructs an update that encodes the backdoor in a way such that, when it is aggregated with other agents' updates, the aggregated model exhibits the backdoor. 

Several works try to prevent such attacks by designing robust aggregation functions~\cite{ yin2018byzantinerobust,pillutla2019robust,blanchard2017machine,mhamdi2018hidden,ozdayi2020defending,sun2019really,fgold,bernstein2018signsgd}.  
In our work, we approach this problem from a complementary angle, considering the fact that some of the proposed defenses are broken~\cite{arxiv:2018:fedlens,arxiv:2018:backdoor}, and there is no guarantee that existing defenses will succeed in defending against any type of adversary. Concretely, we design a framework that incorporates \emph{accountability} to the FL framework to discourage attackers. That is, \textit{even if an attack is not prevented during training, our framework allows one to detect and penalize the adversarial agents later at a time when the backdoor is found in the trained model}.

To make FL accountable, we leverage \emph{blockchains} since they are compatible with the decentralized nature of FL, provide practical immutability, and Turing-complete computation on the logged data via smart contracts. The challenge in this context is to design a blockchain architecture with a low communication and latency overhead such that it can be seamlessly incorporated to the FL data flow. We address this problem by designing a hybrid architecture consisting of both a public, and a private blockchain. This is because latency in public blockchains is too high to run any computation intensive algorithms, and the data on the public blockchains can be accessed by anyone. On the other hand, even though private blockchains are communication efficient, and address privacy challenges by allowing sensitive data to be seen only by an approved set of participants, they do not allow for public accountability since transactions are approved by a predetermined set of users, and cannot be accessed publicly.
Thus, by combining public, and private blockchains in a novel architecture, we alleviate the weakness of each, and have an architecture that meets the needs of FL (cf. Figure~\ref{fig:main_framework} for an overview of our framework).

To our knowledge, this is the first work that implements FL over a hybrid blockchain architecture to discourage attacks by providing accountability, and penalty mechanisms. We also note that, our framework is general in the sense that, any aggregation method, such as FedAvg~\cite{fed-average},  signSGD~\cite{bernstein2018signsgd}, and any attacker detection mechanism can be plugged into it.

\subsection{Overview of Our Contributions}
The key contributions of our work are as follows:
\begin{itemize}
\item We propose \blockfl: \textbf{Block}chain-based \textbf{F}ederated \textbf{L}earning with \textbf{A}ccountablity. \blockfl is a general FL framework that aims to deter adversarial attacks by providing accountability. \blockfl is general in the sense that, any aggregation function, and any attacker detection algorithm can be plugged into it.
\item We provide a novel attacker detection algorithm for FL setting, particularly designed against pixel-pattern backdoor attacks and show its effectiveness empirically.
\item We show extensive empirical evaluation of the \blockfl on different settings.
\item We analyze the security, and privacy provided by \blockfl in detail.
\end{itemize}


The remainder of the paper is structured as follows: In Section~\ref{sec:background},
we provide the necessary background to the reader.  Section \ref{sec:architecture} discusses the system architecture which illustrates a detailed outline of the on-chain aggregation, verification technique over the public chain, the penalty structure, log scheme and finally the trojan detection mechanism. In section \ref{sec:implementation} and section \ref{sec:optimization}, we go over our implementation and some optimization techniques used to improve the performance of the system. In section \ref{sec:experiment} and section \ref{sec:security} we provide experimental evaluation results and subsequently analyze the security and privacy parameters of the system. Section \ref{sec:related} details the comparison of the \blockfl system with other related blockchain based systems that integrate with Federated Learning. We conclude the paper with section \ref{sec:conclusion} while providing some scope for the future work we intend to accomplish.

\section{Background} \label{sec:background}
In this section, we provide the necessary background to the reader.


\subsection{Federated Learning}
At a high level, FL is a multi-round machine learning protocol between an aggregation server and a set of agents, in which agents jointly train a model. Formally, participating agents try to minimize the average of their loss functions,
$$
\argmin_{w \in R^d} f(w) = \frac{1}{K}\sum_{k=1}^K f_k(w),
$$
where $f_k$ is the loss function of k\ts{th} agent. For example, for neural networks, $f_k$ is typically empirical risk minimization under a loss function $L$ such as cross-entropy, \ie,

$$
f_k(w) = \frac{1}{n_k} \sum_{j=1}^{n_k} L(x_j, y_j; w),
$$
with $n_k$ being the total number of samples in agent's dataset and $(x_j,y_j)$ being the j\ts{th} sample.

Concretely, FL protocol is executed as follows: at round $t$, server samples a subset of agents $S_t$, and sends them $w_t$, the model weights for the current round. Upon receiving $w_t$,  k\ts{th} agent initializes his model with the received weight, and train for some number of iterations, \eg, via stochastic gradient descent (SGD), and ends up with weights $w_t^k$. The agent then computes his update as $\Delta_t^k = w_t^k - w_t$ and sends it back to the server. Upon receiving the update of every agent in $S_t$, server computes the weights for the next round by aggregating the updates with an aggregation function $\mathbf{g} \colon R^{|S_t| \times d } \rightarrow R^d$ and adding the result to $w_t$. That is,
$w_{t+1} = w_t + \eta\cdot\g( \{ \Delta_t \} )$ where $\{ \Delta_t \} = \bigcup_{k \in S_t}\Delta_t^k$, and $\eta$ is the server's learning rate.

For example, original FL paper~\cite{fed-learning:google} and many subsequent papers on FL~\cite{arxiv:2018:fedlens,arxiv:2018:backdoor,sun2019really,bonawitz2016practical,geyer2017differentially} consider weighted averaging to aggregate updates. In this context, this aggregation is referred as Federated Averaging (FedAvg), and yields the following update rule,
\begin{equation}
w_{t+1} = w_t + \eta \frac{\sum_{k \in S_t} n_k \cdot \Delta_t^k}{\sum_{k \in S_t} n_k}.
\label{eqn:fedavg}
\end{equation}

Another prominent aggregation method is presented in~\cite{bernstein2018signsgd}. In this work, authors develop a communication efficient, distributed SGD protocol in which agents only communicate the signs of their gradients. In this case, server aggregates the received signs and returns the sign of aggregation to the agents who locally update their models using it. We refer their aggregation technique as \emph{sign aggregation}, and in FL setting, it yields the following update rule,
\begin{equation}
w_{t+1} = w_t + \eta \big( \sign \sum_{k \in S_t} \sign(\Delta_t^k)  \big),
\label{eqn:sign_agg}
\end{equation}
where $\sign$ is the element-wise sign operation.

In practice, rounds in FL can go on indefinitely, as new agents can keep joining the protocol, or until the model reaches some desired performance  metric (\eg, accuracy) on a validation dataset maintained by the server.

\subsection{Backdoor Attacks and Model Poisoning}
Training time attacks against machine learning models can roughly be classified into two categories: targeted~\cite{arxiv:2018:fedlens,arxiv:2018:backdoor, chen2017targeted,liu2017trojaning}, and untargeted attacks~\cite{blanchard2017machine,bernstein2018signsgd}.
In untargeted attacks, the adversarial task is to make the model converge to a sub-optimal minima, or to make the model completely diverge. Such attacks have been also referred as \emph{convergence attacks}, and to some extend, they are easily detectable by observing the model's accuracy on a validation data. 

On the other hand, in targeted attacks, adversary wants the model to misclassify only a set of chosen samples while minimally affecting its performance on the main task. Such targeted attacks are also known as \emph{backdoor attacks}. A prominent way of carrying backdoor attacks is through \emph{trojans}~\cite{chen2017targeted,liu2017trojaning}. A trojan is a carefully crafted pattern that is leveraged to cause the desired misclassification. 
For example, consider a classification task over cars and planes and let the adversarial task be making the model classify blue cars as planes. Then, adversary could craft a brand logo, put it on \emph{some} of the blue car samples in the training dataset, and only mislabel those as plane. Then, potentially, model would learn to classify blue cars with the brand logo as plane. At the inference time, adversary can present a blue car sample with the logo to the model to activate the backdoor. Ideally, since the model would behave correctly on blue cars that do not have the trojan, it would not be easy to detect the backdoor on a clean validation dataset.

In FL, the training data is decentralized and the aggregation server is only exposed to model updates. Given that, backdoor attacks are typically carried by constructing malicious updates. That is, adversary tries to create an update that encodes the backdoor in a way such that, when it is aggregated with other updates, the aggregated model exhibits the backdoor. This has been referred as \emph{model poisoning} attack  \cite{arxiv:2018:fedlens,arxiv:2018:backdoor,sun2019really}. For example, an adversary could control some of the participating agents in a FL instance, and train their local models on trojaned datasets to construct malicious updates.

\subsection{Blockchain}
Blockchain was first introduced by Nakamato as the underlying ledger of the now famous Bitcoin cryptocurrency~\cite{Bitcoin}. Briefly, a blockchain is an append-only, distributed and replicated database. It allows the participants of a network to collectively maintain a sequence of data in a tamper-resilient way. More importantly, it does so without a requirement for a trusted third party by invoking a consensus mechanism.

Informally, a blockchain network operates as follows: participants broadcast their data, and certain nodes called \emph{miners} (or \emph{validators}) gather, and store the data they receive in wrapper structures called \emph{blocks}. Through a consensus mechanism, the network elects a leader miner in a decentralized fashion for a sequence of epochs. The epoch leader broadcast his block to the network and, having received the leaders block, other nodes store it in their local memory where each block maintains a hash-link to the previous block. 

The consensus algorithm that the blockchain network deploys may depend on whether or not the network is \emph{public}.
For example, Bitcoin operates on a public network, where anyone is free to join and there is no uniform view of the network across participants. It utilizes a cryptographic puzzle called Proof-of-Work~\cite{PoW} to achieve consensus. This makes tampering with the order of blocks computationally infeasible when the majority of the network participants follow the protocol honestly. In \emph{private} networks however, participants can employ more efficient consensus algorithms, such as PBFT~\cite{PBFT}. This is because the identity and number of participants are known to every party, as access to the such networks can be arbitrarily restricted.

We provide examples for a private, and a public blockchain below, and note that there exists also hybrid architectures (as in this work), that combine both public, and private blockchains.

\subsubsection{Private Blockchain: Hyperledger Fabric}
Hyperledger~\cite{hyperledger} is the umbrella project for many open source blockchains. Hyperledger Fabric, a permissioned blockchain is one amongst many blockchains that holds properties like identifiable participants, high transaction throughput performance~\cite{8500557}, low latency of transaction~\cite{8038517} confirmation alongside privacy and confidentiality of transactions. Hyperledger promotes the usage of smart contracts called chaincode and pluggable consensus models for the confirmation of the underlying transactions committed on the ledger. The transaction orders are maintained and are visible to all peers participating on the network. 

\subsubsection{Public Blockchain: Ethereum}
Ethereum~\cite{wood2014ethereum}, also possess the capability to host smart contracts. However, the smart contracts published are public due to the permissionless nature of the blockchain making every transaction transparent. Each ethereum smart contract and participant have an account of its own. Ether, being the hosted cryptocurrency on the ethereum chain is required to publish contracts, call functions and send transactions over the chain. This currency is stored in a wallet possessed by every participant on the blockchain and is spent in the form of Gas to make smart contract calls. Ethereum, however, offers low transaction throughput and high latency on transaction confirmation.

\section{System Architecture} \label{sec:architecture}

In this paper, we propose a practical system architecture that allows any Federated Learning  
algorithm to run efficiently and securely while enabling auditability. Our solution maintains a multi-factor approach to securely detect the potential trojan introduced in the model over time and penalize the offending parties. There are many components to the system, each playing a critical role to accomplish the comprehensive goal. 

\subsection{Framework Setup}
The overall \blockfl framework assumes each participant trains the model on their local machine or on a separate Virtual Machine in the cloud. This assumption eliminates the expense of training the model on the chain and enhances data privacy. Alongside training the model locally, we consider the network to be an established TCP connection between the participants and the aggregation server, thus eliminating the overhead for establishing a connection every time an event happens. 
Furthermore, the number of parameters sent by each mini-batch gradient epoch is an arbitrary number. 
There is no trusted server to perform any operations locally. All operations pertaining to inter-blockchain transactions are performed from any node that hosts the private blockchain and that could be either a worker node or an endorser node.

\subsection{On-Chain Aggregation} \label{sec:on-chain-aggregation}
In order to attain a true distributed nature of the Federated Learning algorithms, we allow the training of the model to happen off the chain by individual worker nodes participating in the convergence process. 
The primary advantage to execute the model training process off the chain is that individual worker nodes can provision their own computing power to dispatch the deep learning parameters to the server. We treat each worker node as individual entities sharing the common interest to get rewarded for training the model.  

The server in the Federated Learning setting is represented by the Private Blockchain that performs the aggregation and sends the global updates back to the worker.
As mentioned earlier, we perform all worker-blockchain communication with the help of a smart contract deployed on the private blockchain. The private blockchain is hosted on the same network as each worker node. Since this is a private blockchain, each account is issued by an authority. 
In this case, the node that sets up the private blockchain behaves as the ultimate authority providing membership to each worker node. Certificates are issued which are signed by the sever Certificate Authority (CA) for the worker nodes to send and receive parameters from the server by maintaining a TLS connection over SSL. 
The worker will then wait post upload of the parameter to the private blockchain until it receives a response from the chaincode in the form of aggregated parameters.
On collection of the aggregated parameters, the worker  moves to the next iteration to retrain its model and resend the updated parameters. 

The server being the on chain entity will wait until each worker node has sent the model parameters. Once an update is received from every worker node, the server/private blockchain will confirm the integrity of each instance of the parameters received. The main factors checked would include the size of the update, the type of the update received and whether the sender has used the appropriate credentials to send the update. 

This ends up in the server/private blockchain to conduct formation of the aggregated parameter and the blockchain will send the update to each worker node, triggering an event on the worker node to initiate the next iteration using the updated aggregate. 

\begin{figure}[b]
    \centering
    \includegraphics[width=0.5\textwidth]{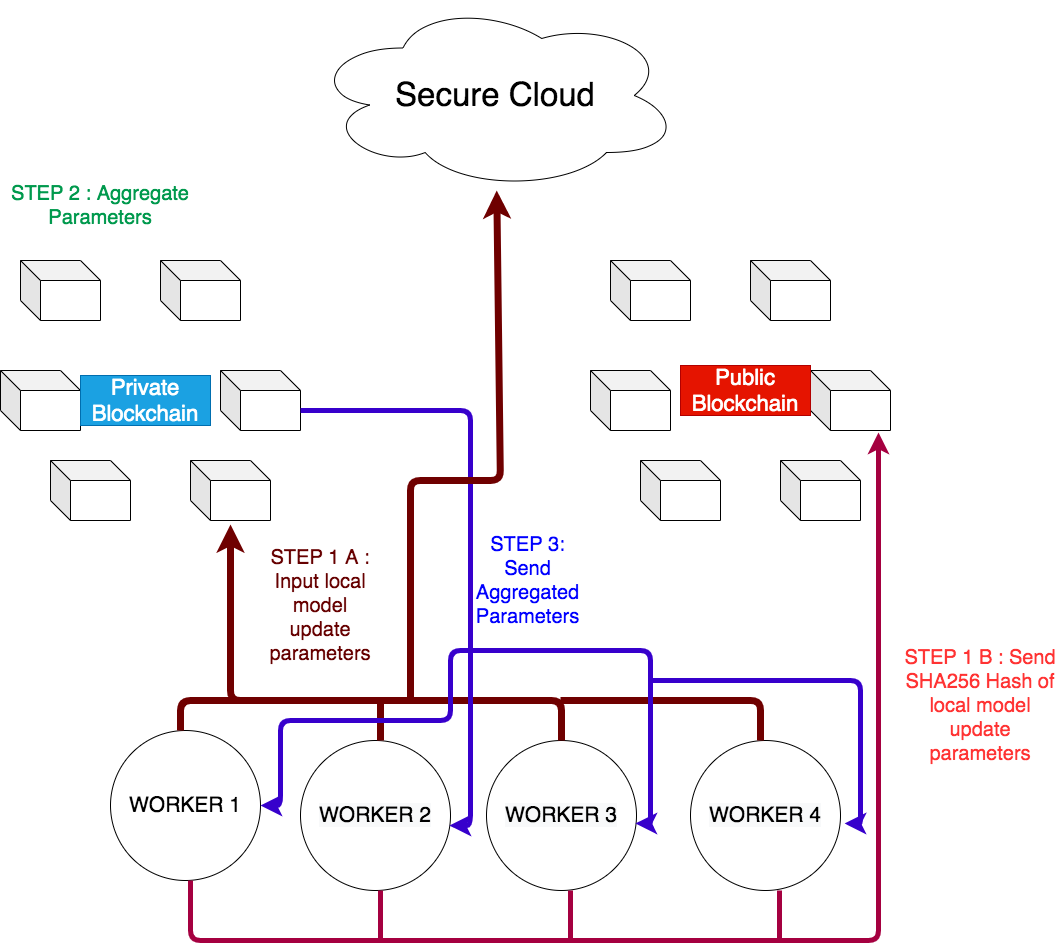}
    \caption{{\bf Federated Learning over Blockchain: Architecture Diagram } 
    }
    \label{fig:main_framework}
\end{figure}

The system architecture illustrated in Figure \ref{fig:main_framework} shows 4 worker nodes as an example participating in the Federated Learning process and generating updates in the form of parameters locally. Step 1A involves sending the local updates/parameters  to the private chain and are also being logged to the secure cloud simultaneously. Logging of local updates to the secure cloud is discussed in section \ref{sec:log-scheme}. Step 1B includes sending the corresponding generated SHA256 hashes of the local updates to the public blockchain for verification purposes and the motivation for this is discussed in section \ref{sec:public-verification}. Aggregation of all the local updates are being done in the private blockchain to generate a global model. Finally in Step 3, the global update in the form of aggregated parameters are being sent back to each worker node for the next epoch.

\subsection{Verification over Public Ledger} \label{sec:public-verification}
The other crucial component of the \blockfl system is the involvement of the public blockchain. Our verification system over the public blockchain is in the form of a smart contract visible to all. 
Every worker node on the private chain has an account on the public chain. The private blockchain maintains a one-to-one mapping of accounts on the private chain to their corresponding accounts on the public chain. Each worker must possess a wallet on the public chain with sufficient crypto-currency or transaction money to call the smart contract functions when sending updates. 

The purpose of the smart contract deployed on the public ledger is three fold.
\begin{enumerate}[label=\alph*)]
	\item Storage of SHA256 Hashes of sent parameters
    \item Verification function to validate whether a participant or a subset of participants have cheated in any fashion.
    \item Deposit and penalty orchestration
\end{enumerate}

The smart contract hosts an array per worker node to store the cryptographic (i.e., SHA256) hashes of \textit{each parameter set} sent to the private blockchain based server. 
The hash value is calculated individually by each worker node off the chain on their own local machine and sent to the public smart contract as shown in Step 1B of Figure \ref{fig:main_framework}. The primary purpose served here is for any worker to hold any offending worker nodes accountable in the case a trojan is detected. As the public chain is transparent and immutable, the worker node can merely download the SHA256, retrieve a recreated SHA256 from the parameter store on the private chain and verify if the SHA256 stored on the public chain matches the private chain created SHA256.

\begin{figure}[b]
     \centering
     \includegraphics[width=0.5\textwidth]{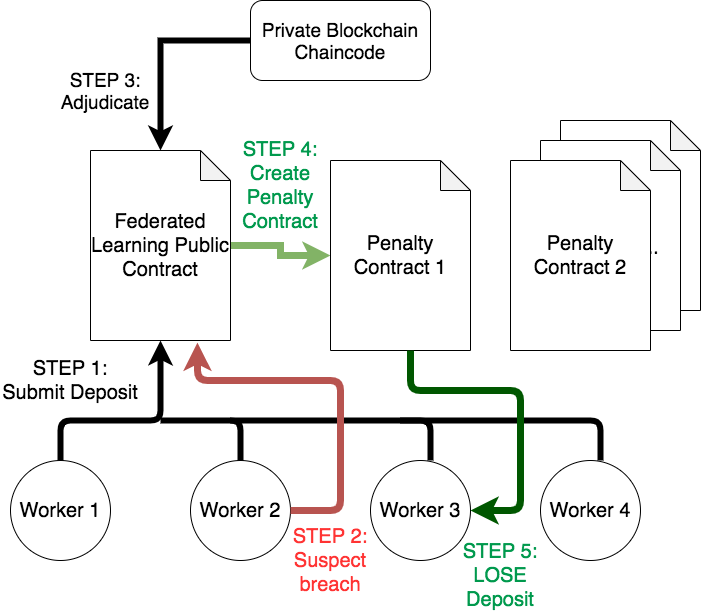}
     \caption{{\bf Breach Adjudications and Penalty Concept of Operations } 
     }
     \label{fig:signSGD_Penalty}
 \end{figure}

As shown in Figure \ref{fig:signSGD_Penalty}, If in case, the SHA256 does not match, then a penalty is administered based on the penalty structure illustrated in the following section.

\subsection{Penalty structure} \label{sec:penalty}
The main incentive for worker nodes for participating as a model trainer in any Federated Learning algorithm is monetary \cite{8726492}. This monetary benefit is fulfilled in the form of crypto-currency. For the purpose of participating as a worker in the Federated Learning process, each node must possess a crypto wallet and submit a deposit over the public chain. This deposit is returned to the worker at completion when convergence of the model is attained. An award is also dispatched to each worker for participating honestly. This award is coupled with a variable reimbursement to each worker that depends on the average time taken to send the model parameters to the server for aggregation. The faster the parameters are sent, the greater the award. 

Although, if a worker is caught sending an update to introduce trojan by any other worker node as discussed in section \ref{sec:anomaly-alert-policy}, the deposit is lost and redistributed to the remaining worker nodes.  For that reason, every worker node also has the opportunity to raise an alarm if a breach has occurred. On suspecting a breach over the public chain, a new verification contract is created and checks whether the accused actually has breached the terms of contract as shown in Figure \ref{fig:signSGD_Penalty}. If the breach is confirmed, then the suspect loses their deposit. If no violation is confirmed, the accuser loses their deposit. This prevents unnecessary breach violation claims which may lock funds on the public chain indefinitely.

\subsection{Log Scheme in Secure Cloud} \label{sec:log-scheme}

Logs are necessary to ensure / reinforce trojan detection. 
There are two important aspects with which the logging mechanism may be able to catch the perpetrator early in the process. 
\subsubsection{Log Publication}
Logs are published to the secure cloud after every epoch by each worker node as shown in Figure \ref{fig:main_framework}. 
The logs are stored in a filesystem based on a secure cloud infrastructure where each worker has its own unique location to publish and after every epoch, the location will have the test classification results from the aggregated parameters excluding that worker node. The logs include the following data:
\begin{enumerate}[label=\alph*)]
	\item Uploaded parameters of each worker node
	\item Test classification results at each epoch after aggregating parameters by excluding the parameters uploaded by that worker. 
\end{enumerate}

\subsubsection{Anomaly alert policy} \label{sec:anomaly-alert-policy}
Whenever a participant suspects a breach, he/she will suspect by calling the public smart contract that holds the SHA256 Hashes of every local update made by every worker. Depending on the input of confirmation from the private blockchain, if the private blockchain is able to generate the SHA256 hashes from the local updates that match the one on the public smart contract, then the suspect was honest in uploading the parameters. 

\emph{Now, for trojan detection, the private blockchain will provide a one-time link access to the location of the logs for the suspect on the secure cloud. The participant will be allowed to download the logs that contain the local parameter updates of the accused party and run the trojan detection algorithm as discussed in section \ref{sec:attack_detect_alg}. If the trojan is detected, then the suspect will be penalized according to the penalty structure as discussed in section \ref{sec:penalty}.}


\subsection{An Attacker Detection Algorithm for Trojaning Attacks}\label{sec:attack_detect_alg}
We now describe a sample attacker detection algorithm, particularly designed for detecting adversaries who run pixel-pattern backdoor attacks, \ie, trojaning (cf. Figure~\ref{fig:trojaned_samples}). However, recall that, \textit{ our framework allows a developer to plug any kind of detection algorithm, or even multiple detection algorithms against different types of attacks, into his system}. \emph{We developed this algorithm for sake of completeness, and to highlight what kind of detection algorithms can be used with our framework.}
Recall that, in trojaning attacks, the adversarial task is to make the model misclassify instances from a \emph{base class} as \emph{target class} by using trojan patterns. To do so, an adversarial agent can simply corrupt his dataset by adding the trojan pattern to his base class instances, relabel them as the target class, and construct malicious updates by training on the corrupted dataset.

In what follows, we assume that the honest participants (i.e., verifiers) has access to \emph{the backdoored model}, \emph{the trojan pattern}, and the information of \emph{base and the target class} used by the adversary. Also verifiers have access to the updates of agents that are sent during the execution of FL, as the updates are logged in the secure cloud (cf. Figure~\ref{fig:main_framework}).

The key idea behind our detection algorithm is to do \emph{parameter attribution} by computing the empirical Fisher Information Matrix (FIM) as done ~\cite{shoham2019overcoming,ozdayi2020defending}. At a high level, a verifier first creates a poisoned validation set using the trojan pattern of the adversary. That is, he extracts base class instances from a clean validation dataset, adds them the trojan pattern and relabels them as the target class. He then replays the training process using the logged updates. After each round, he computes the backdoor loss on the poisoned validation dataset using the aggregated model. Then, for rounds in which backdoor loss decreases, the verifier does parameter attribution via FIM on the aggregated model using the poisoned validation dataset. This allows the verifier to list the parameters of the model in order of importance for the backdoor task, so she can identify the top-$\kappa$ most important parameters for the backdoor task (where $\kappa$ is a hyperparameter). Finally, the developer then measures and records the $L_2$ norm of each agent's update for these $\kappa$ parameters.

The intuition is, \emph{when backdoor loss decreases, we would expect the attackers contribution to be larger than contribution of honest agents for the most important backdoor task parameters.} Then, by looking at the average of recorded $L_2$ norms over the rounds, and making an assumption on the number of adversarial agents, the verifier can attempt to distinguish attackers. That is, the agents that contribute most on the top-$\kappa$ important parameters for the backdoor task are likely to be adversarial. We illustrate the performance of our algorithm in Section~\ref{sec:attack_detect_alg} via experiments.

\section{Implementation} \label{sec:implementation}

Hyperledger Fabric is a permissioned blockchain infrastructure, providing a modular architecture with a delineation of roles between the nodes in the infrastructure, execution of Smart Contracts (called "chaincode" in Fabric) and configurable consensus and membership services \cite{hyperledger}.  Therefore, for our implementation, the choice of private blockchain is Hyperledger fabric. The fabric infrastructure is set up on docker containers hosted on a virtual machine inside Amazon Web Services(AWS). 

We have one top level organization in the fabric implementation that is controlled by the node that sets up the private blockchain. The organization acts as the membership service provider to issue client certificates to the participants. The chaincode written in golang incorporates the logic to accept deep learning(DL) parameters in the form of batches and store them in a key-value store on the ledger. The data structure for each worker stores the deep learning parameters in a json-based tree structure. For example, in order to store the DL parameters of worker A, we divide the parameters into N parts and store them on the chain with keys A01, A02, $\dots$ AN etc.  

Ethereum is one the most popular public blockchain, being the second largest cryptocurrency features a smart contract functionality formed around the principle of consensus thus eliminating the possibilities of fraud, corruption and makes the network tamper-proof. Therefore, the choice of public blockchain is Ethereum Ropsten \cite{kim2018measuring}. We use go-ethereum node to deploy our solidity based smart contract on the public chain. Our wallet is controlled by metamask and each participant has an account inside the wallet. The server deploys the contract making it the owner. The SHA256 hashes on the smart contract is stored in the form of arrays. Each worker has their own array and each element of the array constitutes the SHA256 hash of the DL parameters uploaded to the private chain in a particular epoch. 

We set up the hyperledger fabric framework on EBS(Elastic Block Storage) backed m5.12xlarge EC2(Elastic Cloud Compute) instances. The specification of the EC2 instance is 24 cores, 192 GiB RAM and a network bandwidth of 10 Gbps. The number of threads per core being 2, the number of vCPUs allotted for processing amounts to 48.

We set up parallel loading of parameters into the private chain, and parallel aggregation of parameters on the chaincode. The SHA256 converter is set up locally on all the worker nodes.


In this work, we implemented two of the most commonly used federated learning averaging techniques: 1) \emph{signSGD} 2) \emph{FedAvg} (see section \ref{sec:background} for more details).
The client application that computes the parameters for the \emph{signSGD} or \emph{FedAvg} alogrithm is implemented in Javascript with the help of Mxnet library. We create nodejs subroutines on the worker nodes to upload DL parameters in batches to the chaincode. A counter is implemented in the chaincode to increment its value whenever an the chaincode accepts the parameters from all the participants. No participant can upload the parameters twice in the same epoch. Once the counter's value becomes equal to the number of participants, the aggregation function is triggered calculating the aggregation based on all the participants' inputs. 
Another nodejs subroutine is implemented to upload the SHA256 hashes generated by the converter to the Ropsten.

In the case of \emph{signSGD}, each worker node trains their model locally and sending the sign of each parameter in binary form  to the smart contract. For the sake of simplicity, we map a negative sign to the bit 0 and the positive sign to bit 1. In the case of \emph{FedAvg}, each worker node trains their model locally and sends the 32 bit float/real number of each parameter in float32 form  to the smart contract.

For \emph{signSGD} again, the Hyperledger Fabric chain code based server then computes the majority bit received from all the workers at each i\textsuperscript{th} position of the parameter. In the case of \emph{FedAvg}, the chaincode computes the average of the parameters received from all the workers at each i\textsuperscript{th} parameter.

\section{Optimization Techniques} \label{sec:optimization}

Maximizing throughput and minimizing communication overhead necessitates for the following optimization strategies we have implemented for a more reliable, practical and efficient execution of the averaging algorithm used for FL over blockchain to reach algorithmic convergence faster. 
To achieve this goal, we created Nodejs subroutines to convert binary parameters to base64 parameters as a compression technique. Also, we implemented a variable thread spawning mechanism to create new chaincode to perform the upload and aggregation of parameters in parallel for optimization purposes. We discuss the details of these optimizations below.

\subsection{Binary compression using base64}
Hyperledger Fabric does not accept inputs in a binary format when workers are sending binary based models in signSGD. The binary parameters have to be sent to the fabric chaincode in a text format. Therefore to send binary input as is, the DL parameters must be sent as  characters in a string.

For example, in signSGD, for model updates involving a DL model that has 300 million parameters, if the binary representation is converted to byte format and sent to the chaincode, we send 37.5 million bytes in the form of 75 million characters achieving a compression of 75\%. 
This however could be improved by compressing the byte representation of the parameters into a base64 representation and decoding the base64 characters into binary on the chaincode. We use golang provided base64 decoding functions to decode the base64 inputted text into binary format. Base64 encodes 3 bytes (6 characters in a string) on 4 characters. Therefore, 37.5 million bytes (75 million characters) can be sent as 50 million base64 characters thus achieving a cumulative compression of 83.33\%. 

With this compression, we send fewer bytes to the Hyperledger Fabric chaincode, thus reducing communication and improving efficiency for signSGD.

\subsection{Preventing key collisions}

Hyperledger Fabric's underlying database uses Multiversion Concurrency Control(MVCC) to guarantee no double spending or inconsistency in the data occur. Therefore, an attempt to update the same state will result in a new version of the existing state being created to overwrite the old one. 

Due to the size of the parameters being uploaded into the hyperledger fabric chaincode, we divide the input into smaller batches to avoid any throughput related errors. Hence, there is a need to parallelize sending input to the Fabric chaincode. If the uploader is not parallelized, then sequential sending of the batches will result in frequent update of the Hyperledger state. Therefore, roughly half of the transactions executed during the upload phase fail due to an MVCC related error 
and the Federated Learning algorithm implementation will be in an inconsistent state.

Hyperledger Fabric currently has the limitation of not being able to handle MVCC conflicts and will allow the transaction to happen resulting in an error being thrown when it tries to execute. Therefore, we design our chaincode data model so that MVCC conflicts are avoided. We separate the read operations from write operations so that we can implement them as queries and invokes respectively. We modified our data model so that every new transaction writes to a completely different key, thus mitigating the problem only to a certain extent. However, we have to record deltas and as a result our updates over key A (A being a participant) would be stored by transactions in independent keys such as A01, A02, A03. The updates implemented in this approach then get aggregated and the current state of A is reflected in the next query. 

\subsection{Architectural Improvements}

\subsubsection{Increase Endorsers and Channels}
Due to usage of cloud based infrastructure and the ability to scale vCPUs, we increased the number of channels and endorsers to increase transaction throughput. We deployed multiple chaincodes on multiple channels to accept and aggregate parameters in parallel. Merely increasing the number of endorser peers do not improve efficiency. The peers are CPU-intensive and as a result have to be placed on separate VMs. 

\subsubsection{Usage of LevelDB}
Hyperledger Fabric deploys with LevelDB in its default setting. However, in order to have a highly available setting, CouchDB is used to maintain the state database. Although, there are benefits in using CouchDB to store binary data modeled in a chaincode and having CouchDB as an external database for resiliency, the performance of LevelDB implementation of the chaincode is significantly better than the CouchDB implementation. LevelDB is a key-value store and practical implementations of it demonstrate significant performance improvement. 

\subsubsection{Deploying Peers on separate nodes}
In Hyperledger Fabric, endorser peers are CPU intensive processes. If deployed in large numbers, they may significantly affect the transaction throughput performance. Orderers and Membership Service Providers / Organizations on the other hand are non-CPU intensive processes. Thus, the peers are deployed on separate Virtual Machines instead of all on one to improve the throughput. We set network bandwidth on the order of 10 Gbps in order for the transaction throughput to remain high. 

\subsection{Cache Chaincode}

Cache chaincode are solely designed to terminate MVCC conflicts to avoid impairing the performance of the network. On uploading inputs to the chaincode, if an MVCC conflict occurs, the parameters can be cached onto this cache chaincode, so that it can be re-uploaded when the execution is complete. 
That way, if the database goes into an inconsistent state due to an error, the cache chaincode will be used to restore the state to maintain consistency. Thus, the worker only has to upload the parameters once and rely on the chaincode architecture to maintain persistence.

\begin{figure*}[t]
    \centering
    \begin{subfigure}[t]{0.5\textwidth}
        \centering
        \includegraphics[width=1\textwidth] {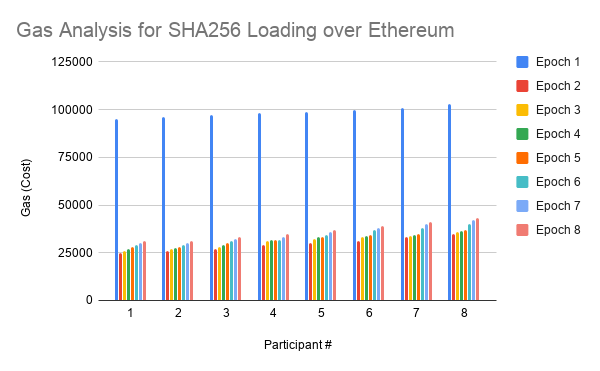}
        \caption{{\bf Gas Analysis for SHA256 Loading over Ethereum } }
        \label{fig:gasAnalysis}
    \end{subfigure}%
    \begin{subfigure}[t]{0.5\textwidth}
        \centering
        \includegraphics[width=1\textwidth] {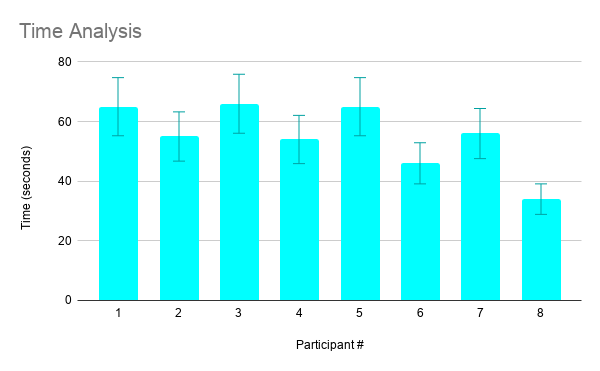}
        \caption{{\bf Time Analysis for SHA256 Loading over Ethereum }}
        \label{fig:timeAnalysis}
    \end{subfigure}%
    \caption{\emph{Experimental results of Cost Analysis over Public Blockchain Ethereum.} }
\end{figure*}

\section{Experimental Evaluation} \label{sec:experiment}

To assess the the \blockfl system, we experiment multiple facets of the system under different scenarios. Our experimentation is pertinent to testing the impact of integrating a hybrid blockchain setting with the Federated Averaging algorithm and SignSGD algorithm, evaluating the penalty structure and security assessment of the system. We examine the effects of parallelism, deploying multiple contracts, using disparate data structures, swapping the underlying blockchain and fiscal impacts on implementing the system.


\subsection{Experiment Setup}

We use hyperledger provided fabric docker containers to be deployed on the m5.12xlarge EC2 instance. We have metamask installed on our browsers with the connection network pointing to the Ropsten Test Network. Our Ethereum version of smart contract is published to Ropsten at address:

0x304095B3Af015DB179CADD70dF4BC9D1fDc1aDbc 

We use openssl to generate self-signed Certificate Authorities (CA) and issue client certificates to the peer nodes signed by the CAs. Nodejs scripts are written to control the deployment of chaincodes and creation of channels in a command-driven routine. A nodejs script is implemented to act as the liaison between the hyperledger fabric and ethereum networks. We use etheruem smart contract auto-generation technique for penalty payments on the public chain. Appropriate security groups are implemented for the endorsing peer based EC2 instances  to communicate with each other. Each account on the Ropsten Ethereum chain are given 10 Ether to begin with and are requested from the Ropsten Ethereum faucet available on the internet. Binary to base64 converter is implemented for signSGD to convert participant provided parameters into base64 to upload to the chaincode. We use the peer binary provided by Hyperledger Fabric to upload parameters in a repeatable fashion.

Since our framework implements FL algorithms as it is, \textit{our implementation does not result in any difference with respect to the accuracy of the generic FL models}. Hence we do not report any accuracy results here. On the other hand, at each epoch, depending on the DL model learned and the averaging scheme used, the performance may greatly change due to blockchain based smart contract characteristics. To understand the factors that impact the overall performance, we experimented with a single round of update under different aggregation schemes with varying DL model sizes. In order to set up our experiments, we adjusted the DL model sizes (i.e., the number of trainable parameters used by DL model) based on the commonly used DL models \cite{the-link-sent-by-mustafa}. For example, classic DL architecture such as AlexNet has around 60 million trainable parameters. On the other hand, newer architectures such as VGG16 has around 160 million trainable parameters. In our experiments, we measure the performance by varying the number of parameters updated at each epoch ranging from 50 million to 300 million. This range covers most of the commonly used DL models \cite{the-link-sent-by-mustafa}.


\begin{table*}
  \caption{Deploy and Aggregation Time Analysis over Contract Variation}
  \label{tab:contract}
  \begin{subtable}[t]{0.49\textwidth}
  \centering
   \begin{tabular}[width=1\textwidth]{ccl}
    \toprule
    \# Contracts& Contract Deploy& Parameter Aggregate\\Deployed& Time& Time\\
    \midrule
    1 & 45 seconds & 4 seconds\\
    2 & 71 seconds & 3.5 seconds\\
    4 & 93 seconds & 3.5 seconds\\
    8 & 144 seconds & 3 seconds\\
    16 & 243 seconds & 2.8 seconds\\
    32 & 443 seconds & 2.5 seconds\\
    48 & 503 seconds & 2.3 seconds\\
  \bottomrule
\end{tabular}
\caption{signSGD}
\label{tab:signSGD}
\end{subtable}
\begin{subtable}[t]{0.49\textwidth}
\flushright
  \centering
  \begin{tabular}[width=1\textwidth]{ccl}
    \toprule
    \# Contracts& Contract Deploy& Parameter Aggregate\\Deployed& Time& Time\\ 
    \midrule
    1 & 35 seconds & 6.6 seconds \\
    2 & 48 seconds & 6.5 seconds \\
    4 & 70 seconds & 6.5 seconds\\
    8 & 132 seconds & 7.3 seconds\\
    16 & 258 seconds & 6.8 seconds\\
    32 & 420 seconds & 6.5 seconds\\
    48 & 510 seconds & 6.5 seconds\\
  \bottomrule
\end{tabular}
\caption{FedAvg}
\label{tab:FedAvg}
  \end{subtable}
\end{table*}


\subsection{Results}

\begin{figure}[b]
    \centering
    \includegraphics[width=0.5\textwidth]{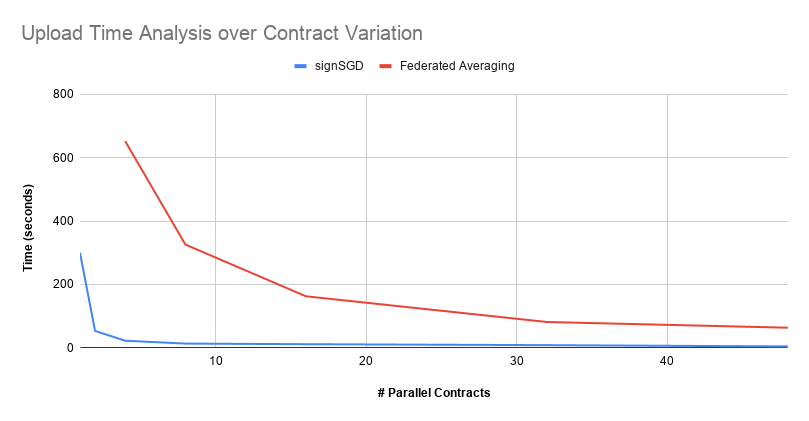}
    \caption{{\bf Upload Time Analysis over Contract Variation } 
    }
    \label{fig:results_contractVariation}
\end{figure}

%

\begin{figure}[b]
    \centering
    \includegraphics[width=0.5\textwidth]{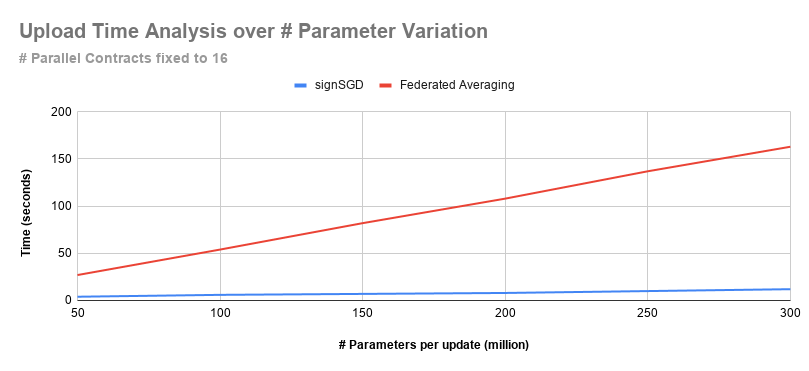}
    \caption{{\bf Upload Time Analysis over Parameter Variation }} 
    \label{fig:results_parameterVariation}
\end{figure}



\subsubsection{1 contract vs N contracts (SignSGD)}

Each client loads upto 10 KB (20,000 characters) of parameters in every iteration of uploading the  DL parameters per epoch. That means per function call the clients are loading up to 13,300 base64 characters. With both, our local and AWS setup, we have compared the cases of using a single contract versus multiple contracts to upload the parameters and subsequently perform aggregation. Parameter loading to the hyperledger fabric framework is done in parallel in case there are multiple contracts. Each series of uploads to the smart contract will have its own thread running in the background. 
For a single thread / single contract system, the parameters are loaded in at the maximum 350 transactions with an event update post each update to give the client application the notification to send another update. 
Since the parameters are uploaded sequentially, the amount of time taken to upload the parameters is up to 5 minutes in both the local setup and AWS setup. This seems to be a reasonable time considering that we are using only a single contract for uploading the parameters sequentially. There is no monetary/gas cost analysis provided since it is a private blockchain and there are no wallets in a private blockchain. We then increased the number of contracts by a factor of 2 and compared the time taken for uploading the parameters to the framework. With our setup, we were able to upload all parameters within 60 seconds.

On further increasing the number of contracts, the time to upload improved significantly. With 32 parallel contracts, we were able to achieve upload time of 9 seconds for any signSGD setting. And if we push the system to its limits by uploading all the parameters in 48 parallel contracts, then the upload time is 5 seconds.  This is the primary advantage of using a private blockchain instead of a public one simply because the transaction throughput is so high. Having achieved 20,000 transactions per second \cite{gorenflo2019fastfabric}, in the event we have better processing power, we can potentially upload all the parameters within a fraction of a second.

The system we were using was a 48 vCPU system. As a result, even if we use more than 48 threads to upload all the parameters, it will have a negligible impact on the performance simply because all the vCPUs have been taken up by the 48 threads earlier. 

The aggregation time as seen in Table \ref{tab:signSGD}, for any Federated Learning signSGD setting is always less than 5 seconds. The results shown in Table \ref{tab:signSGD} are recorded to demonstrate the worst case timing. Realistically, the parameters sent by each worker node would not be above 300 million parameters. Therefore, we have shown that even in the worst case scenario, aggregation happens within 5 seconds.

Thus, if we add up the results seen in Table \ref{tab:signSGD} and Figure \ref{fig:results_contractVariation}, for each worker node, with the current system, for 5-10 epochs, the total time taken for the aggregation and parameter loading to run takes less than 1.5 minute. 

\subsubsection{1 contract vs N contracts (FedAvg)}
Each client loads 130 KB (33000 float32) in every iteration of uploading all Federated Averaging parameters per epoch. 
We use the same experimental setup used for \emph{signSGD}. For a sequential upload of parameters, the amount of time taken to upload the parameters is 25 minutes in both the local setup and AWS setup. This drastic increase in time is again, due to the usage of events, that means we will not make another upload until the previous upload is complete.

On gradually increasing the number of contracts as seen in Figure \ref{fig:results_contractVariation}, the time to upload improved significantly. With 16 parallel contracts, we were able to achieve upload time of under 3 minutes for any Federated Averaging setting. And if we push the system to its limits by uploading all 330 million parameters in 48 parallel contracts, then the upload time is under 1.5 minutes. Being a private blockchain instead of a public one implies the transaction throughput is so high.

The aggregation time as seen in Table \ref{tab:FedAvg}, for any Federated Averaging setting is less than 10 seconds constantly because it is being done on the ledger and will consume only one transaction to return the results. Similar to Table \ref{tab:signSGD}, the results shown in Table \ref{tab:FedAvg} demonstrate the worst case timing for aggregating 300 million parameters.

Thus, if we add up the results seen in Table \ref{tab:FedAvg} and Figure \ref{fig:results_contractVariation}, for each worker node, with the current system, for 5-10 epochs, the total time taken for the complete \emph{FedAvg} algorithm to run takes less than 12 minutes. 

\subsubsection{Performance Analysis on Parameter Variation}
As shown in Figure \ref{fig:results_parameterVariation}, on increasing the number of parameters to be uploaded by each Federated Learning algorithm by keeping the number of parallel contracts deployed constant(her 16 parallel contracts), the time required to upload all parameters into the private blockchain increases gradually. For signSGD, it takes about 4 seconds to upload 50 million parameters to the private chain and 12 seconds to upload 300 million parameters to the private chain. Similarly, for Federated Averaging, it takes 27 seconds to upload 50 million parameters to the private chain and 163 seconds to upload 300 million parameters to the public chain. This tells us that with the current optimizations in place, we are able to achieve communication efficient Federated Learning over the Hybrid Blockchain setting.

\subsubsection{Hybrid chain vs all Public chain}

We also deployed a parallel solidity smart contract that performs the same operations as the golang hyperledger chaincode and deployed it on the ethereum private blockchain. There was a drastic slowness observed when performing Federated Averaging operations over Ethereum. The throughput attained when inputting parameters to the ethereum smart contract 
more than 1300 seconds for \emph{FedAvg} versus 
64 seconds per 300 million parameters for \emph{FedAvg} for the hyperledger fabric framework.  
The time taken to perform aggregation was also of the order of 450 seconds for ethereum versus less than 10 seconds for hyperledger fabric. Since a private deployment of the Ethereum blockchain was set up, there are no incurred transaction costs incurred to any worker node. 

\subsubsection{Gas and Time Analysis over Public Chain}
The price of 1 Ethereum Gas is 0.02 $\mu$ Ether. The amount of gas it takes to upload the SHA256 Hashes to the Ropsten Ethereum public chain is nearly 95000 Gas (.0019 Ether) for the first call of the first epoch. On subsequent epochs, each call takes nearly 25000 Gas(0.0005 Ether) to upload the SHA256 Hashes as shown in Figure \ref{fig:gasAnalysis}. Cost to upload the Hash value is more for the first call because it initializes the data structure on the smart contract, thereby being more expensive than merely updating and appending values to the array that is already created. As seen in Figure \ref{fig:timeAnalysis}, the SHA256 Hash upload from the client is variant between 35 seconds to 75 seconds. This means each transaction confirmation over the public chain is independent of size of the hash and therefore unreliable to store the real parameters on the public chain if performance is taken into consideration.

\subsection{Performance of our Attacker Detection Algorithm}
\label{sec:detection_res}
\begin{figure}[h]
\includegraphics[scale=0.30]{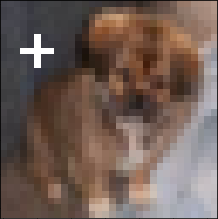}
\includegraphics[scale=0.30]{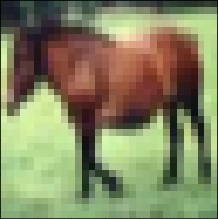}
 \caption{Samples from trojaned base class, and target class.
 The trojan pattern is a 5-by-5 plus pattern that is put to the top-left of base class instances. The goal of adversary is to make the model classify instances of dog class with the trojan pattern (left) as a horse (right).}
\label{fig:trojaned_samples}
\end{figure}

We now illustrate the performance of the attacker detection algorithm we described in Section~\ref{sec:attack_detect_alg} via experiments. The general setting of our experiments are as follows: we simulate FL for $R$ rounds among $K$ agents where $F$ fraction of them are corrupt. The backdoor task is to make the model misclassify instances from a \emph{base class} as \emph{target class} by using trojan patterns. That is, a model having the backdoor classifies instances from base class with trojan pattern as target class (see Figure \ref{fig:trojaned_samples}). To do so, we assume an adversary who corrupts the local datasets of corrupt agents by adding a trojan pattern to base class instances, and relabeling them as target class. Other than that, adversary cannot view and modify updates of honest agents, or cannot influence the computation done by honest agents and the aggregation server. At each round, the server uniformly samples $C\cdot K$ agents for training where $C \leq 1$. Those agents locally train for $E$ epochs with a batch size of $B$ before sending their updates.

We tested our detection algorithm on CIFAR10~\cite{cifar10} dataset by using a 5-layer convolutional neural network consisting of about 1.2M parameters with the following architecture: two layers of convolution, followed by a layer of max-pooling, followed by two fully-connected layers with dropout. Hyperparameters used in all experiments can be found in Appendix~\ref{app:hyperparams}. We were able to detect the adversarial agents perfectly in three out of these four settings. We briefly summarize our results below.

\begin{figure}[h]
\centering
\includegraphics[scale=0.5]{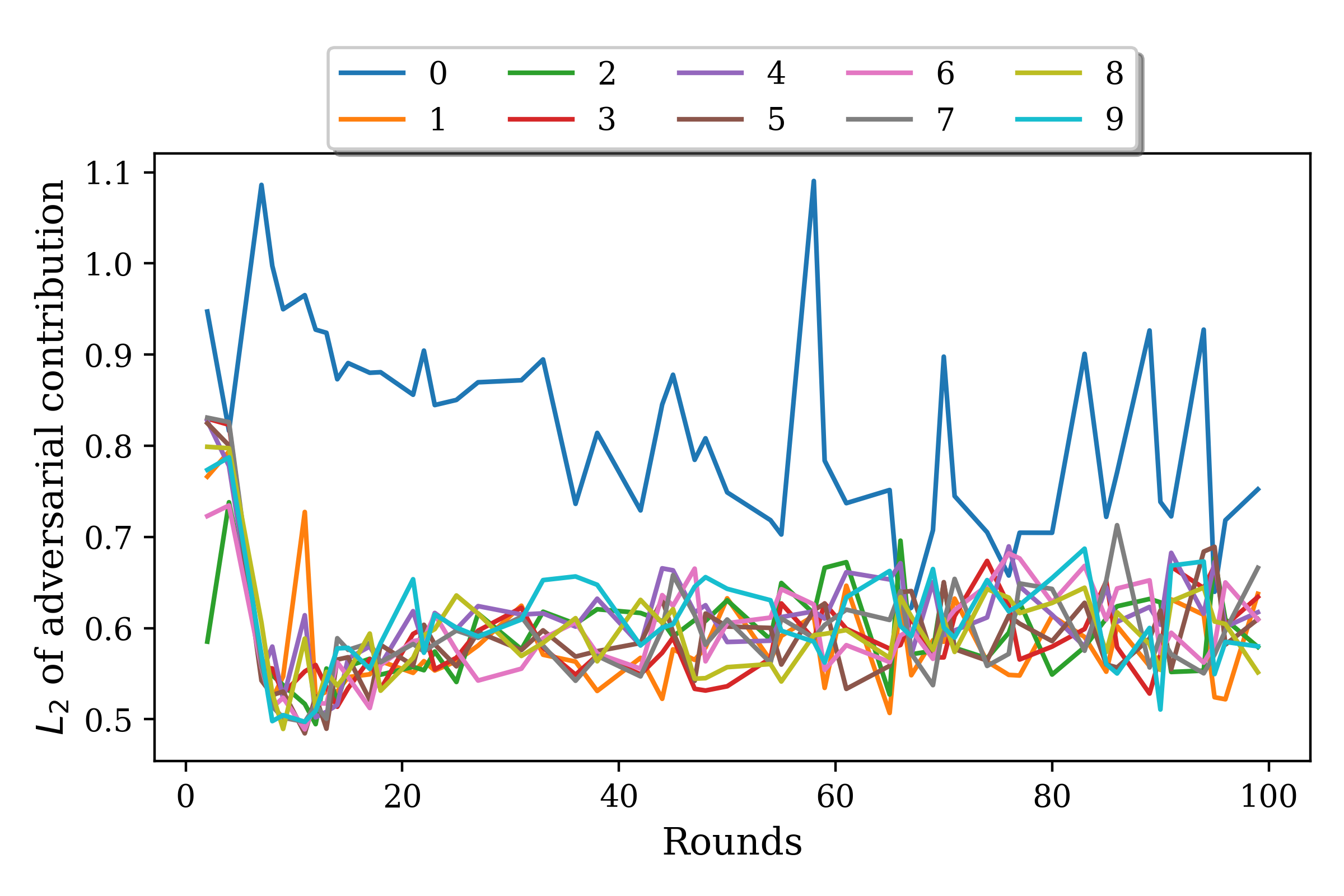}
\includegraphics[scale=0.5]{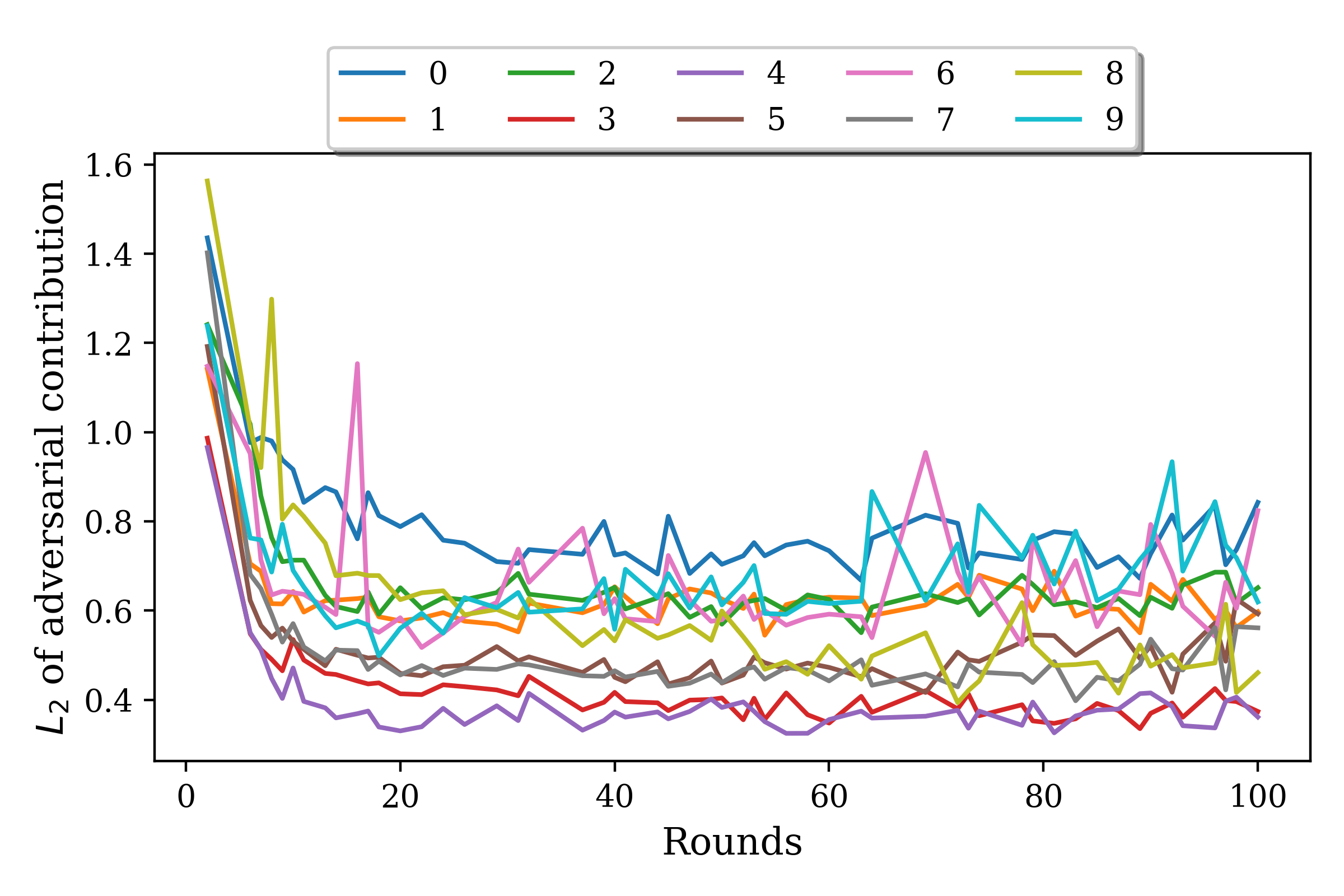}
 \caption{Results of trojan detection algorithm in iid (top), and non-iid (bottom) settings. In iid setting, the adversarial agent (Agent 0) stands out very clearly. In non-iid setting, the variance of contribution of agents' seem to be higher. Yet, if we compute the average $L_2$ contribution over the rounds, Agent 0 is at the top with a value of $0.78$,  followed by Agent 9 ($0.67$) and Agent 6 ($0.65$). In iid setting, data is distributed uniformly between agents, and in non-iid setting, we distribute each class by sampling from a Dirichlet distribution with a concentration of $0.5$}
\label{fig:detection_small}
\end{figure}

In the first two settings, we have a rather small setup of 10 agents, where one of them is adversary. We plot the $L_2$ of adversarial contribution for the rounds in which backdoor loss decreases in Figure~\ref{fig:detection_small}. The attacker is Agent 0, and as can be seen, he stands out from the rest by having the largest $L_2$ contribution to most important backdoor parameters.

Our last two settings are somewhat more realistic for FL. We have a setup of 100 agents, where 10 of them are corrupt, and where in each round, the aggregation server uniformly samples 10 agent out of 100. Our detection algorithm were able to perfectly distinguish adversarial agents for iid case. That is, \textit{when we list agents by adversarial contribution, the top-10 consisted of solely adversarial agents}. However, for non-iid setting, it included some non-adversarial agents. For that case, in top-5, we had 4 adversarial agents, but the places between 6-10 were non-adversarial agents.
We suspect this is because the variance of adversarial contributions are relatively higher in larger, and non-iid settings.
Again,\emph{we stress that our framework could be used with other detection algorithms}.

\section{Security and Privacy Analysis} \label{sec:security}

In this section, we provide an overview of the security and privacy assurances of the proposed system. Especially, we discuss how the \blockfl system handles attacks from the adversaries discussed in section 2.4.

\subsubsection{Security Analysis}
Since aggregation done on the private blockchain, as long as the 51\% of the participants are honest (needed for convergence of the consensus algorithm used), the averaging needed for the FL will be correctly executed. Therefore, the aggregation will be done according to the specified protocol. Therefore, any single attacker cannot change the aggregation process.

On the other hand, a malicious worker node can try to attack the system by introducing a backdoor /trojan. Assuming the detection technique work with high probability,  the participants may be held accountable for an attack. 
Since all participants commit the SHA256 Hashes of their parametric updates to the public smart contract, we can easily detect incorrect disclosure of updates. If for any reason, the participant introduces a backdoor, then the federated learning algorithm can be re-simulated on the private chain using the updates as the test data using the appropriate off-chain detection technique.


\subsubsection{Privacy Analysis}
The privacy protection of parameters sent over the private blockchain is achieved with the help of channels. Each worker is connected to the server by a separate channel thus maintaining strict access control on the parametric data being shared between the worker node and the server over the private chain. Since each participant can join the hyperledger framework only with  authorization from the membership service provider, the data parameter updates sent between the worker nodes and the server node remain on the chain concealed from those who have no account on the private chain. Moreover, for the participant to access their own data, TLS (Transport Layer Security) over the private blockchain is used for granting access to the data. Thus, each participant's  privacy over the public blockchain is achieved with the help of hashing (i.e., single hash of  millions of model parameters shared at each round). An external user or adversary can only retrieve the SHA256 hash of the parameters being exchanged over the private chain. Log files over the public cloud is encrypted and accessible only by the server. 

\section{Related Work} \label{sec:related}
A privacy-preserving solution for federated learning has been proposed in BlockFlow \cite{mugunthan2020blockflow}, they consider differential privacy utilizing Laplacian noise to avoid disclosing client information to other other clients. Still, this does not address the possibility of detection of an anomaly based trojan being introduced during the learning process. 
Moreover, BlockFlow performs aggregation and evaluation of client models over the ethereum blockchain which is a public chain. 
However, implementing both the aggregation and penalty mechanism over the public chain drastically slows down the federated learning process. Moreover, sending models over the public chain results in expensive cost in terms of gas/ether incurred by both the client and the server which is not beneficial to the algorithm. 
This issue is addressed by our system. Since federated learning is eminently reliant on communication between the worker and server nodes, we have proposed and implemented a hybrid blockchain architecture to perform the aggregation over the private chain and the penalty mechanism over the public chain. Also having a separate smart contract between each client and server by the auto contract generation will keep the penalty mechanism more structured and robust.

The BC-FL framework proposed in \cite{ma2020federated} discusses the integration of a public blockchain with federated learning to prevent malicious worker nodes or UEs (User Equipment) as they call it from poisoning the learning process. They identify the issue that federated learning aggregation being conducted on a central server is a single point of failure in the federated learning process. The agenda of their system primarily involves model recording and publishing over the public chain and an incentive mechanism to motivate miners and discourage lazy nodes. To improve security while maintaining efficiency, they propose a Digital Signature system to recognize malicious clients by verifying learning results and a reputation system to reduce verification delay of a high-reputation miner. Although, the proposal makes an attempt, the system itself discourages efficiency since the federated learning aggregation is being conducted solely on the public chain. As we have seen in  \cite{kone2017federated}, Federated learning without a communication-efficient framework will take a considerable amount of time to reach convergence. Public blockchains like ethereum are inherently slow to commit transactions and this will cause a significant impact to Federated Learning performance. The \blockfl framework, however, addresses this issue by implementing a hybrid blockchain framework. Secondly, the Digital Signature and the reputation system only prevents external attackers from introducing bad models. For internal attackers, it relies on input from the rest of the participants. Moreover, if a backdoor is introduced, it is not possible to be detected by the participants themselves. As we discussed, we can detect a backdoor introduced with the help of our framework.

\section{Conclusions and Future Work} \label{sec:conclusion}
Federated Learning is an upcoming communication-efficient machine learning technique that trains a global model while maintaining data local. However, due to an existing risk of backdoor attacks, the poisoning of the global model can yield ineffective results in terms of classification and prediction which defeats the purpose of the overall goal of FL.

In order to address this aforementioned problem, we have proposed a general blockchain framework that integrates both public and private blockchains to achieve decentralization, immutability and transparency to discourage backdoor attacks on federated learning algorithms by holding the responsible parties accountable. We have shown that our framework facilitates the adaptation to any federated learning algorithm rendering a plug-and-play setting. We have presented the implementation of the Federated Averaging and signSGD algorithm 
over our general blockchain framework and based on our empirical results, conclude that our proposed approach maintains the communication-efficient nature of these algorithms. 

We have also proposed and implemented an algorithm that runs off the chain server-side to successfully detect the attacker who introduced the trojan for the Federated Averaging algorithm. Our experiments show that we are able to effectively detect the participant who has introduced the model poisoning attack. Furthermore, the blockchain-based penalty system proves to be an efficient deterrent for an attacker to carry out any model poisoning attacks by injecting backdoors/trojan. 

As a future work, we plan to develop a trojan detection algorithm for signSGD.  This seems to be challenging due to the communication involving only bits instead of real numbers. Still, we believe this is feasible by using more trojan examples. In addition, we will look into integrating other federated learning aggregation algorithms in our system.

\clearpage



\bibliographystyle{ACM-Reference-Format}
\bibliography{references}

\appendix
\section{Hyperparameters of Experiments}\label{app:hyperparams}
We remind the notation we used for our experiments, and report the hyperparameters accordingly. 
\begin{itemize}
    \item{R: Number of rounds }
    \item{K: Total number of agents}
    \item{F: Fraction of corrupt agents}
    \item{C: Fraction of selected agents for training in a round}
    \item{E: Number of epochs in local training} 
    \item{B: Batch size of local training}
    \item{$\eta$: Server's learning rate}
    \item{$\kappa$: Number of parameter to compute $L_2$ norm for (cf. Section~\ref{sec:attack_detect_alg}.)}
\end{itemize}

\begin{table}[h!]
\centering
\label{tab:hyperparamsIID}
\resizebox{0.7\columnwidth}{!}{%
\begin{tabular}{  c c c c c c c c c c } 
 \toprule
 R & K & F & C & E & B & $\eta$ & $\kappa$ \\
 \midrule
 100 & 10 & 0.1 & 1 & 2 & 256 & 1 & 1000 \\
 \bottomrule
 \end{tabular}}
 \caption{Hyperparameters for small setting.}
 \end{table}
 
\begin{table}[h!]
\centering
\label{tab:hyperparamsIID}
\resizebox{0.7\columnwidth}{!}{%
\begin{tabular}{  c c c c c c c c c c } 
 \toprule
 R & K & F & C & E & B & $\eta$ & $\kappa$ \\
 \midrule
 200 & 100 & 0.1 & 0.1 & 5 & 32 & 1 & 1000 \\
 \bottomrule
 \end{tabular}}
 \caption{Hyperparameters for large setting.}
 \end{table}
\end{document}